\documentclass[
    amsmath,
    amssymb,
    superscriptaddress,
    aps,
    prx,
    reprint,
    floatfix,
]{revtex4-2}

\usepackage{graphicx}
\usepackage{dcolumn}
\usepackage{glossaries}
\usepackage[version=4]{mhchem}
\usepackage[per-mode=symbol]{siunitx}
\usepackage[most]{tcolorbox}
\usepackage{xcolor}
\usepackage{hyperref}
\usepackage{threeparttable}
\usepackage{booktabs}

\hypersetup{
    unicode,
    colorlinks=true,
    urlcolor=black,
    linkcolor=black,
    citecolor=black
}

\DeclareMathOperator{\sech}{sech}

\newacronym{bec}{BEC}{Born effective charge}
\newacronym{dft}{DFT}{density functional theory}
\newacronym{dw}{DW}{domain wall}
\newacronym{irrep}{irrep}{irreducible representation}
\newacronym{md}{MD}{molecular dynamics}
\newacronym{mlip}{MLIP}{machine-learned interatomic potential}
\newacronym{neb}{NEB}{nudged elastic band}
\newacronym{nep}{NEP}{neuroevolution potential}
\newacronym{pppm}{PPPM}{particle-particle particle-mesh}
\newacronym{qnep}{qNEP}{charge-aware NEP}
\newacronym{rmse}{RMSE}{root-mean-square error}
\newacronym{rp}{RP}{Ruddlesden-Popper}

\makeatletter
\let\oldtheequation\theequation
\renewcommand\tagform@[1]{\maketag@@@{\ignorespaces#1\unskip\@@italiccorr}}
\renewcommand\theequation{(\oldtheequation)}
\makeatother

\DeclareSIUnit\angstrom{\text{Å}}
\DeclareSIUnit\site{\text{site}}
\DeclareSIUnit\atom{\text{atom}}
\DeclareSIUnit\rad{\text{rad}}
\newcommand{\doHMN}[2]{%
  \begingroup\lccode`~=`#1
  \lowercase{\endgroup\let~}#2%
  \mathcode`#1="8000
}


\newcommand{\ntnu}{
    Department of Materials Science and Engineering, Faculty of Natural Sciences and Technology,
    NTNU Norwegian University of Science and Technology, NO-7491 Trondheim, Norway 
}
\newcommand{\phys}{
    Department of Physics and Astronomy, 
    Chalmers University of Technology,
    SE-41296 Gothenburg, Sweden
}
\newcommand{\dur}{
    Centre for Materials Physics, Durham University, South Road, Durham, DH1 3LE, United Kingdom
}

\begin{document}
\title{Machine-learning-guided exploration of domain walls \texorpdfstring{\\}{} in the hybrid improper ferroelectric \ce{Ca3Ti2O7}}

\author{Ida C. Skogvoll}
\affiliation{\ntnu}
\author{Erik Fransson}
\affiliation{\phys}
\author{Leo Ö. Westin}
\affiliation{\phys}
\author{Benjamin A. D. Williamson}
\affiliation{\ntnu}
\author{Nicholas C. Bristowe}
\affiliation{\dur}
\author{Sverre M. Selbach}
\email{selbach@ntnu.no}
\affiliation{\ntnu}
\author{Paul Erhart}
\email{erhart@chalmers.se}
\affiliation{\phys}

\hypersetup{pdfauthor={Ida C. Skogvoll, Erik Fransson, Leo Ö. Westin, Benjamin A. D. Williamson, Nicholas C. Bristowe, Sverre M. Selbach, Paul Erhart}}

\begin{abstract}
Ruddlesden-Popper phases are highly tunable and naturally layered structures, in which polarization can arise via a hybrid improper ferroelectric mechanism.
This enables a complex \gls{dw} structure where multiple order parameters, like octahedral rotations, polar distortions and strain, interact. 
In this work, we explore the rich set of  \gls{dw} structures in prototypical \ce{Ca3Ti2O7}, mapping out the \glspl{dw} in the $\{100\}$, $\{110\}$ and $\{001\}$ pseudo-tetragonal planes using group theory and \glspl{mlip}.
The trained potential reproduces the \gls{dft} order parameter and polarization profiles for all wall types and orientations considered.
A charge-aware training framework combined with reference Born effective charges reduces the prediction errors for the \gls{dw} formation energies by \qty{70}{\percent}, revealing the importance of including long-range electrostatics to model symmetry-broken interfaces. 
Finally, the \gls{mlip} is used to identify minimum energy pathways at the atomic scale with nearly the precision of \gls{dft} calculations, revealing a low-energy antipolar configuration for polarization switching.   
\end{abstract}

\maketitle
\glsresetall

Ferroelectric materials, with their spontaneous and switchable polarization, are a crucial component in current and emerging device technologies.
With advances in the spatial resolution of imaging and characterization techniques, these applications now extend beyond bulk polarization switching and include nanoscale device architectures that functionalize the domain boundaries themselves \cite{nataf2020domain}.
These \glspl{dw} are two-dimensional topological defects where the symmetry-breaking and confined structural distortion enable compelling emergent phenomena \cite{meier2015functional, meier2022ferroelectric, seidel2009conduction}.
They are also naturally occurring interfaces that can be dynamically controlled by an external field, as opposed to fixed heterostructures.
Device concepts cover atmospheric sensing, diodes, memory, and circuitry in unconventional computing schemes \cite{mundy2017functional, whyte2015diode, richarz2024ferroelectric, sanchez2017resonant, maksymovych2011dynamic, sun2022memory}. 

At the heart of controlling ferroelectric switching, regardless of the particular device design, is understanding the microscopic properties of the domain structure and \gls{dw} configuration.
This is because polarization switching and hysteretic behavior are fundamentally driven by \gls{dw} motion.
Modeling the dynamic behavior of complex domain structures requires finite temperature simulations on the atomistic or mesoscale level.
Large-scale atomistic simulations in this field, commonly based on \gls{md}, have been mostly confined to the prototypical proper ferroelectrics, like \ce{PbTiO3}, \ce{BiFeO3}, and \ce{BaTiO3} \cite{shin2007nucleation, liu2013exploration, liu2016intrinsic, ovcenavsek2023dynamics, Gomez-Ortiz2023, Khachaturyan2024, sehrawat2025machine}.
In these materials the primary order parameter of the ferroelectric phase transition is a single polar lattice distortion, originating from covalent mechanisms such as lone-pair or second-order Jahn-Teller effects \cite{halasyamani1998noncentrosymmetric}.  

In improper ferroelectrics, on the other hand, the ordering transition is driven by a non-polar lattice distortion that couples indirectly to a secondary polar mode arising, for instance, from geometric constraints \cite{levanyuk1974improper}.
Since the polarization is not the primary order parameter but is locked to a robust structural distortion \cite{benedek2022hybrid}, it is less readily suppressed by the depolarization fields that degrade device performance in thin films \cite{lu2023out}.
A promising class of such materials emerged with the discovery of hybrid improper ferroelectricity \cite{bousquet2008improper, benedek2011hybrid}, characterized by two non-polar lattice distortions coupled to a polar mode through a trilinear coupling mechanism.
In the \gls{rp} phases, which have a quasi-two-dimensional layered perovskite structure, this connection is established by two octahedral rotational distortions breaking inversion symmetry through the activation of a zone-center mode. 

In \ce{Ca3Ti2O7} and its \ce{Sr}-doped counterpart \ce{(Ca,Sr)3Ti2O7}, the composite form of the order parameter results in a complex domain structure of intertwined ferroelastic and ferroelectric \glspl{dw} with varying charge states \cite{huang2016domain, smith2019infrared, gao2017interrelation}.
For bulk and epitaxial thin films of \ce{Ca3Ti2O7}, a switchable polarization of \qty{8}{\micro\coulomb\per\centi\meter\squared} was experimentally verified \cite{oh2015experimental, li2017ultra}.
Several other isostructural \gls{rp} phases of different chemical compositions have also been found to exhibit ferroelectricity, such as \ce{(Ca,Sr)3Sn2O7} and \ce{Sr3Zr2O7} \cite{benedek2022hybrid}. 
Due to the improper nature of the ferroic order, these materials also introduce an avenue for robust magnetoelectric coupling, where the magnetic order and polarization are determined by the same distortion \cite{lu2023out, zhu2024thermal, markovic2020electronically}. 
The \gls{rp} crystal structure is highly tunable through A-site substitution or by controlling the perovskite slab thickness, resulting in a rich phase diagram \cite{chen2025coexistence, kayastha2025diverse}.
Straining has been shown to promote superconductivity \cite{jerzembeck2022superconductivity, sakakibara2024theoretical} while the reduced dimensionality gives favorable band gaps for optoelectronic applications in chalcogenides and halides \cite{zhang2016directed, zhang2017ferroelectricity, li2019band}.

Theoretical studies have made important progress toward understanding how these properties relate to the crystal structure and provide insight into the atomistic mechanisms that underpin ferroelectric switching.
These include thermal and strain evolution of distortion modes during phase transitions \cite{pomiro2020first, clarke2024pressure}, and group-theoretical identification of possible low-energy pathways \cite{nowadnick2016domains, li2020suppressing}.
These calculations are performed using bulk structures and cannot capture the confined geometry, electrostatics, or other gradient-dependent properties of \glspl{dw}. 
There are currently only a few first-principles studies explicitly investigating \glspl{dw} for the \gls{rp} phase \cite{stone2019atomic, nowadnick2016domains}, while large-scale \gls{md} simulations are lacking.
This is likely due to the complexity of the domain structure and large supercells required to resolve the order-parameter switching at the wall.
Such supercells are far beyond what is tractable with \gls{dft}, but are in principle accessible with \glspl{mlip}.
Whether a conventional \gls{mlip} suffices is not obvious, since \glspl{dw} involve a discontinuous polarization and long-range electrostatic contributions that lie outside a local description of the energy \cite{zhong2025machine}.

Here, we address this gap with a comprehensive study of the static and dynamic properties of ferroelastic and ferroelectric \glspl{dw} in the prototypical \gls{rp} \ce{Ca3Ti2O7}.
Group-theoretical analysis is used to map out the symmetry-allowed wall configurations in the $\{100\}_\text{pt}$, $\{110\}_\text{pt}$ and $\{001\}_\text{pt}$ planes, and formation energetics, order parameter, and polarization profiles are characterized using \gls{dft} and \gls{nep}-based \glspl{mlip}. 
Notably, \gls{qnep} models trained with target \glspl{bec} lower the \gls{rmse} by \qty{70}{\percent} compared to a conventional \gls{nep} training scheme, and all local \gls{dw} geometries and polarization profiles are accurately reproduced.
Minimum energy pathways of the $\{100\}_\text{pt}$ walls confirm that the ferroelectric and ferroelastic walls in \ce{Ca3Ti2O7} switch through a polar state, while the purely ferroelectric walls split into an antipolar $Pnma$ \qty{90}{\degree} configuration that lowers the electrostatic energy.
These findings clarify the switching mechanisms in $Cmc2_1$ \gls{rp} phases and can explain the anomalously low coercive field reported for \ce{Ca3Ti2O7} \cite{li2017ultra}. 
Overall, this work demonstrates that \gls{nep} models can accurately and efficiently simulate the structural and electrostatic properties of complex domain structures.

\section*{Results}
\begin{figure*}
    \centering
    \includegraphics[width=\linewidth]{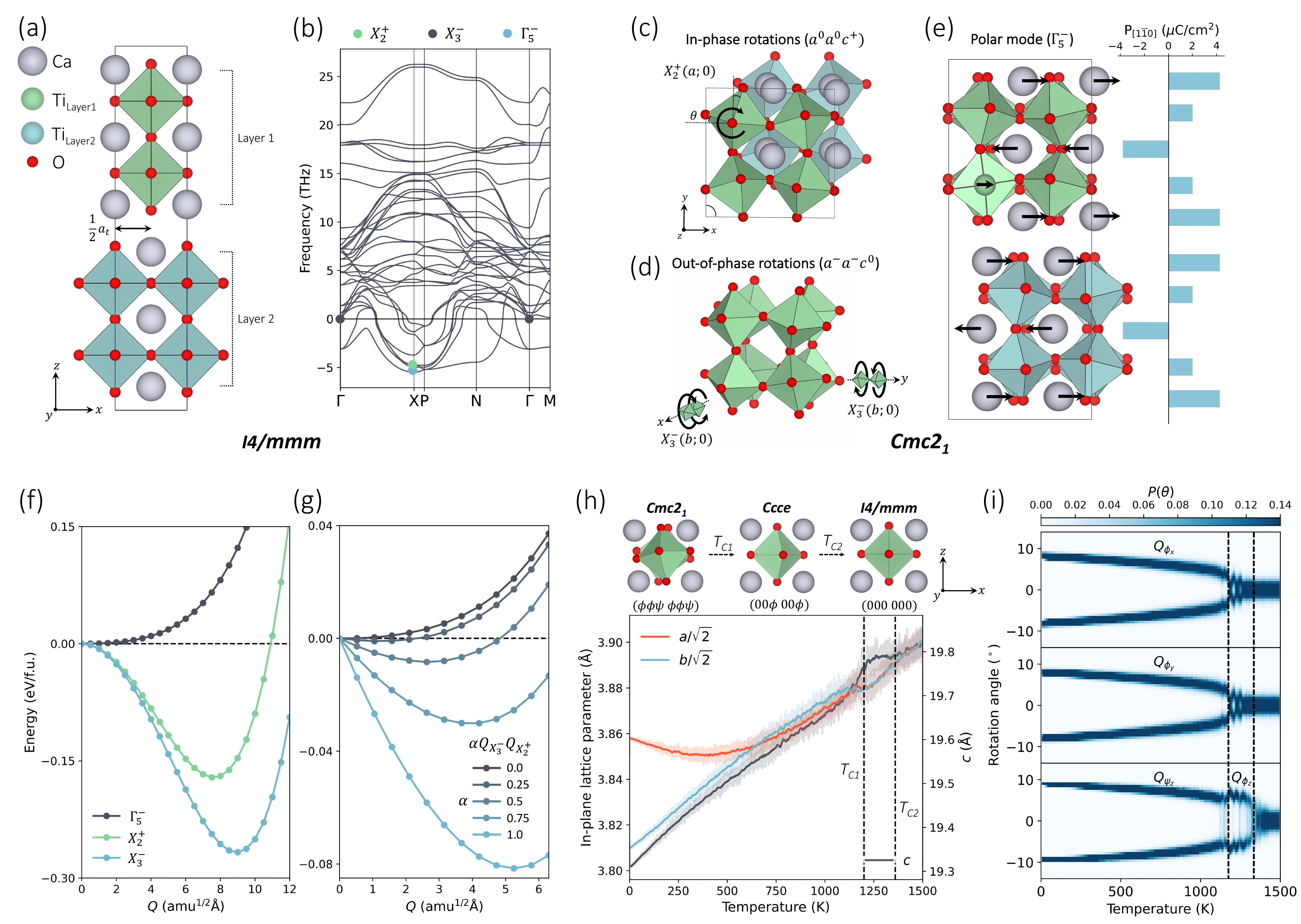}
    \caption{
    (a) The aristotype tetragonal structure of the \gls{rp} phase and the high-temperature phase of \ce{Ca3Ti2O7}.
    Each perovskite slab contains two layers of \ce{TiO6} octahedra, displaced by half of the face diagonal of the $ab$-plane.
    (b) The phonon dispersion of the $I4/mmm$ phase, with markers for distortion modes present in the ground state.
    (c--d) Rotational distortion modes that make up the orthorhombic $Cmc2_1$ phase.
    The $X_2^+$ \gls{irrep} represents in-phase rotations around the out-of-plane axis, while $X_3^-$ are out-of-phase rotations around the $x$ and $y$ axes.
    The twin angle ($\gamma$) is indicated by the arc in the lower left corner,  $90^\circ+\gamma$.
    (e) The polar mode consists of alternating in-plane displacements (black arrows) of the \ce{Ca} atoms, resulting in a single layer of uncompensated \ce{Ca} atoms for all even $n$. 
    There is an off-centering of \ce{Ti} from the octahedron center of mass contributing to the total polarization.
    Note that the unit cell used is an extended cell with respect to the primitive cell. 
    To the right is the calculated polarization (in the $[1\overline{1}0]$ direction) for each atomic layer.
    (f) Energy as a function of mode amplitude ($Q$), when applying the $\Gamma_5^-$, $X_2^+$ and $X_3^-$ modes to the $I4/mmm$ structure. 
    (g) Energy as a function of the polar mode amplitude, $Q(\Gamma_5^-)$, at fixed amplitudes of the rotational modes. $\alpha = 1$ corresponds to the equilibrium displacements from (f).  
    (h) Pseudo-tetragonal lattice parameters and (i) tilt angle evolution when heating to \qty{1500}{\kelvin}. Black dashed lines show the transition temperatures, $T_{C1}$ and $T_{C2}$, separating the $Cmc2_1$, $Ccce$ and $I4/mmm$ phases.
    }
    \label{fig:fig1}
\end{figure*}

The \gls{rp} structure comprises a series of compounds with general formula \ce{A_{n+1}B_nX_{3n+1}}, where $n$ denotes the number of octahedral layers in the quasi two-dimensional \ce{(ABX3)_n} perovskite slabs, each separated by an \ce{AX} rocksalt layer.
The aristotype has space group $I4/mmm$, where the octahedral layers are displaced relative to each other by half the face diagonal of the unit cell (\autoref{fig:fig1}a).
On cooling from this ideal phase, in-phase or out-of-phase octahedral rotations appear, distinguished by whether successive octahedra along a given axis rotate in the same or in opposite directions. 

The ground state has $Cmc2_1$ symmetry with a $\phi\phi\psi_z\:\phi\phi\psi_z$ tilting pattern in Aleksandrov notation \cite{aleksandrov2001structural}.  
$\phi$ denotes out-of-phase rotations and $\psi$ indicates in-phase rotations, and $\psi_z$ is the tilting around the `out-of-plane' direction perpendicular to the octahedral layers.
In the more commonly encountered Glazer notation, this corresponds to an $a^-a^-c^+$ pattern in each slab \cite{stokes2002group}.
The latter notation does not account for the two perovskite slabs being symmetrically distinct, which allows for inequivalent tilting patterns in each layer and the possibility of a relative phase difference in the case that they are identical.
There is for instance a distinction between the polar $Cmc2_1$ and non-polar $Pnma$ ($\phi\phi\psi_z\:\phi\phi\bar{\psi_z}$) phases, although they are energetically very close.
The unit cell is $\sqrt{2}a_\text{t} \times \sqrt{2}a_\text{t} \times c$, $a_\text{t}$ being the in-plane lattice parameter of the tetragonal phase.
From here on all referenced directions will be given in a coordinate basis coinciding with the pseudo-tetragonal lattice vectors and the tilt axes, where $z$ is the out-of-plane direction.

The phonon dispersion of the $I4/mmm$ phase exhibits a range of unstable imaginary modes at multiple high-symmetry points (\autoref{fig:fig1}b).
This is indicative of the flexibility of the \gls{rp} phase for engineering tailored structural or functional properties, and reflects the large number of competing low-energy structures.
At the $X$-point, the first and third lowest phonon modes correspond to the in-phase and out-of-phase rotations found in the $Cmc2_1$ structure, with \glspl{irrep} $X_2^+$ and $X_3^-$, respectively.
These are the primary order parameters in the antiferrodistortive transition (\autoref{fig:fig1}c--d).
The second band at the $X$-point is the $X_1^-$ mode responsible for out-of-phase rotations in the $z$-direction.
The polar distortion in the ground state is a stable and doubly degenerate $\Gamma_5^-$ mode.
It consists of an alternating in-plane displacement of the \ce{Ca} atoms, and the off-centering of \ce{Ti} from the center of the \ce{TiO6} octahedra (\autoref{fig:fig1}e).
The remainder of the imaginary modes correspond to unconventional tilting or cation displacements not found through mode decomposition of the ground state.

Since the polar mode is stable, applying it to the high-symmetry structure gives a single-well potential with no further energy-lowering, as opposed to the double well potentials of the rotational modes, see \autoref{fig:fig1}f. 
However, the rotational modes are non-polar distortions and cannot break inversion symmetry on their own.
Hybrid improper ferroelectricity therefore originates as a secondary effect, through the trilinear coupling mechanism between these rotational modes and the zone-center polar mode.
This manifests itself as a free energy term,
\begin{align}
    F = \alpha Q_{X_3^-}Q_{X_2^+}Q_{\Gamma_5^-},
    \label{eq:free_energy}
\end{align}
where $Q$ is the mode amplitude and $\alpha$ is the coupling coefficient.
As a result, the polar mode becomes energetically favorable only once the rotational distortions are already active,  as shown in \autoref{fig:fig1}g \cite{lei2018observation, benedek2022hybrid}.
This free energy term is analogous to the $R_4^+ \oplus M_3^+ \oplus X_5^+$ coupling between octahedral rotational modes and antipolar A-site distortion in the perovskite structure \cite{mulder2013turning, skogvoll2025local}.
However, because of the termination at each layer in the \gls{rp} structure, there is a net displacement producing an in-plane polarization for each layer (when $n$ is even).

The calculated total polarization of the ground state is \qty{17.5}{\micro\coulomb\per\centi\meter\squared}, close to the structural polarization of \qty{20}{\micro\coulomb\per\centi\meter\squared} obtained by Rietveld refinement \cite{Senn2015}.
It exceeds the largest switchable polarization reported for bulk single crystals, \qty{8}{\micro\coulomb\per\centi\meter\squared} \cite{zhang2021review}, which is expected since the calculation describes a monodomain system at \qty{0}{\kelvin} and therefore excludes both thermal and disorder contributions.
The layer-resolved contributions to the total polarization for successive \ce{CaO} and \ce{TiO2} atomic layers also show that there is a slightly higher polarization associated with the edge \ce{CaO} layers as opposed to the internal \ce{CaO} layer (\autoref{fig:fig1}e).

The sequence of phase transitions in \ce{Ca3Ti2O7} was investigated with \gls{md} simulations using the final \gls{qnep} potential (see Methods). 
It is known that \ce{Ca3Ti2O7} transitions from the ideal tetragonal $I4/mmm$ structure into an intermediate non-polar phase, before reaching its ferroelectric and ferroelastic $Cmc2_1$ ground state.
However, the symmetry of the intermediate phase has been a subject of discussion, with several different space groups proposed \cite{liu2015hybrid, gao2017interrelation, wu2020remarkable}. 
In simulations of the lattice parameter evolution on heating from \qty{1}{\kelvin} to \qty{1500}{\kelvin} (\autoref{fig:fig1}h), the polar ground state transitions into a non-polar orthorhombic phase at \qty{1200}{\kelvin} and into the tetragonal high-symmetry phase at around \qty{1360}{\kelvin}. 

Tracking the octahedral rotation angles (\autoref{fig:fig1}i), the structure maintains the $\phi\phi\psi_z\:\phi\phi\psi_z$ tilt pattern up to $T_{C1}$, where the tilts around the $x$ and $y$ axes vanish and the rotations around the $z$-axis flip from in-phase ($\psi_z$) to out-of-phase ($\phi_z$), yielding the $Ccce$ phase. The remaining $\phi_z$ rotations disappear at $T_{C2}$, completing the $Cmc2_1\rightarrow Ccce \rightarrow I4/mmm$ sequence.
This is consistent with a more recent high-temperature diffraction study \cite{pomiro2020first}, and is analogous to the tilt pattern of the three phases in \ce{CaTiO3} \cite{YASHIMA2009120}. 
We also note that there is a crossover between the $a$ and $b$ lattice parameters close to \qty{600}{\kelvin}, even though the octahedral tilting stays rigid. 
This could be explained by a competition between tilt modes and internal \ce{Ti-O-Ti} angle distortion, as observed in three-dimensional perovskites like \ce{CaMnO3} and \ce{LaFeO3} \cite{SELBACH2012249, zhou2006thermal, DIXON2015337}.
The \gls{qnep} potential was also used to successfully reproduce the phase transitions in \ce{CaTiO3}. 
For more details on the analysis of these simulations, see the Supplementary Information \cite{SM}. 

\subsection*{Domain wall geometry}

\begin{figure*}
    \centering
    \includegraphics[width=1\linewidth]{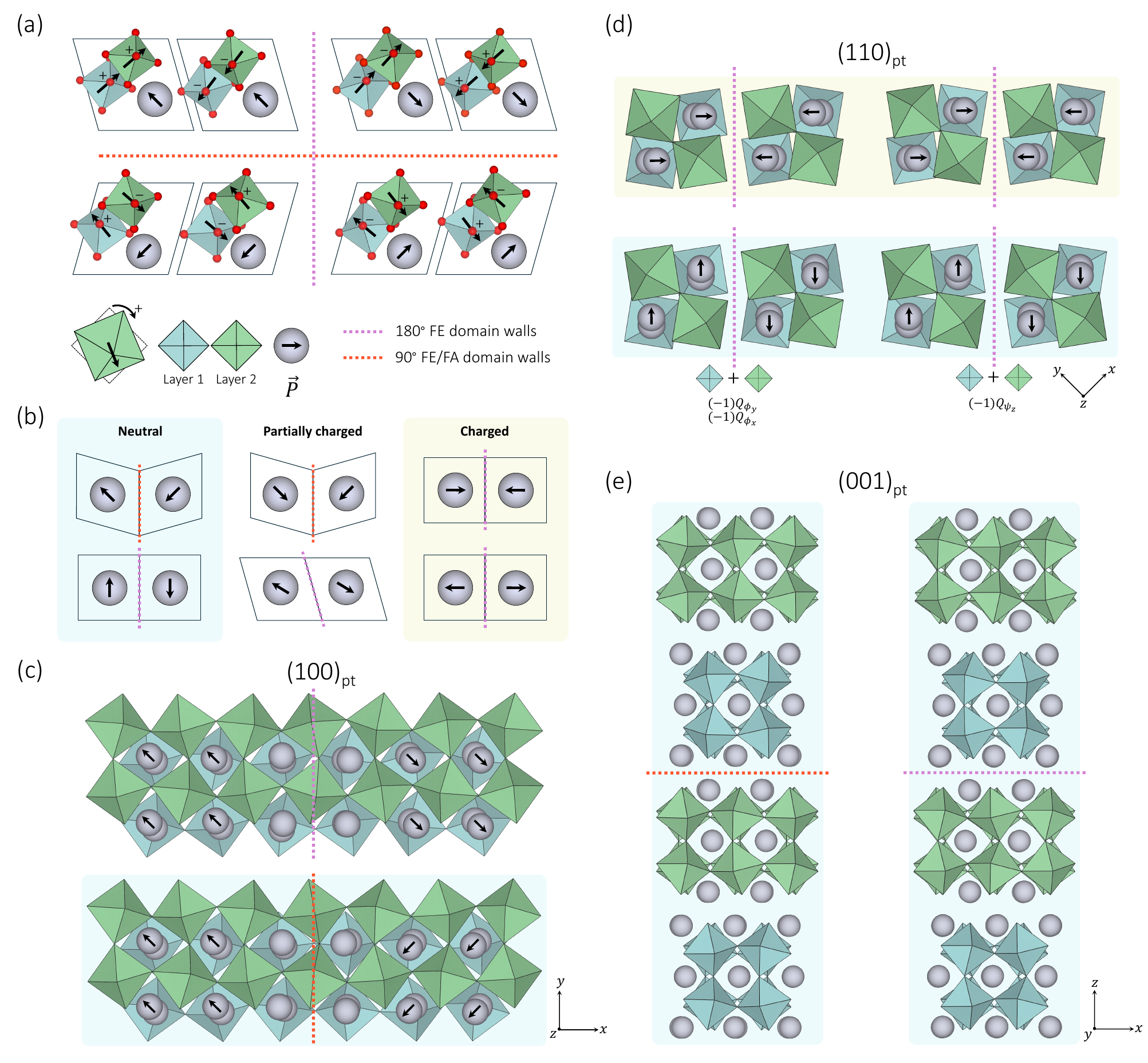}
    \caption{(a) Each of the eight domain states for \ce{Ca3Ti2O7} in the $Cmc2_1$ phase as seen from the $z$-direction, characterized in terms of the tilting of a specific octahedron, the strain state and the polarization ($\vec{P}$) direction.
    For the octahedra, the $+$ and $-$ signs represent the clockwise or counterclockwise rotation of $\psi_z$, while the arrow shows the combined tilting state of $\phi_x$ and $\phi_y$.
    The strain is represented by the cell shape with an exaggerated twin angle ($\gamma$).
    Any combination of domains separated by the purple dashed line creates a \qty{180}{\degree} ferroelectric (FE) \gls{dw}, while the red line separates a \qty{90}{\degree} ferroelectric and ferroelastic (FA) \gls{dw}. (b) The different charge states used to characterize the walls along with examples in the $(100)_\text{pt}$ and $(110)_\text{pt}$ planes.  
    (c) Examples of relaxed $\phi_x\psi_z-\phi_y$ and $\phi_x\phi_y$ $(100)_\text{pt}$ \glspl{dw}.
    (d--e) Examples of charged and neutral \glspl{dw} in the $(110)_\text{pt}$ plane, and \qty{90}{\degree} and \qty{180}{\degree} \glspl{dw} in the $(001)_\text{pt}$ plane.}
    \label{fig:fig2}
\end{figure*}

The antiferrodistortive rotations generate a spontaneous strain in \ce{Ca3Ti2O7} that is switchable under shear stress.
It is therefore a ferroelastic as well as ferroelectric material.
The strained state manifests itself as a twin angle of \qty{0.6}{\degree} in the extended orthorhombic unit cell (\autoref{fig:fig1}c).
The ferroelectric distortion is locked to this strain state through the trilinear coupling, most notably through the off-centering of A-site cations to accommodate the octahedral rotations.
The resulting polarization points in the $\langle 110\rangle$ directions.
Enumeration of the possible configurations for the \glspl{irrep} $X_3^-$ and $X_2^+$ in the $Cmc2_1$ space group yields eight different symmetry equivalent domains.
Each of these can be uniquely categorized in terms of the strain state, tilts and rotations of a single octahedron and the polarization direction (\autoref{fig:fig2}a).

Since a ferroelastic \gls{dw} is defined in terms of a change in strain state, it is  accompanied by a \qty{90}{\degree} rotation of the polarization.
Following the structure of \autoref{fig:fig2}a, combining any domain state on each side of the red dashed line yields a \qty{90}{\degree} wall which is both ferroelastic and ferroelectric. 
Purely ferroelectric \glspl{dw} are associated with a \qty{180}{\degree} rotation of the polarization, resulting from combining domains within each row on either side of the purple dashed line.
The various walls will therefore be referred to here as either \qty{180}{\degree} or \qty{90}{\degree} walls to distinguish between them being purely ferroelectric or also ferroelastic. 
Walls separating domains of equivalent polarization direction but different rotational states (within each quadrant) produce structural antiphase boundaries not considered in this work. 

Although there are multiple ways of combining these domains to form a ferroic wall, there is a small number of symmetry-distinct \glspl{dw}, which can be systematically mapped out by symmetry analysis.
Following Refs.~\citenum{oh2015experimental, huang2016domain}, the trilinear coupling term of \autoref{eq:free_energy} can be decomposed in terms of the tilting around each axis as
\begin{align}
    F_\text{Layer} = \alpha_1 Q_{\phi_x}Q_{\psi_z}P_x + \alpha_2 Q_{\phi_y}Q_{\psi_z}P_y.
\end{align}
$F_\text{Layer}$ applies separately to each of the two layers, giving one such equation per layer, and the tilt amplitudes in the two equations are coupled through space group symmetry.
$\phi_x$ and $\phi_y$ are the out-of-phase rotations around the $x$ and $y$ axes, while $\psi_z$ is the in-phase rotation which is always in the out-of-plane direction.
A \qty{180}{\degree} wall is then associated with switching both $P_x$ and $P_y$, achieved by reversing either $\phi_x$ and $\phi_y$ or $\psi_z$ in both layers independently.
On the other hand, a \qty{90}{\degree} wall is produced by switching either $P_x$ or $P_y$, which is done by reversing the tilts of $\phi_x$ and $\psi_z$ in one layer and $\phi_y$ in the other, and vice versa.
Thus, there is a layer-dependent wall asymmetry for twin walls, and each type is distinguished by what $\phi$-component is perpendicular to the wall plane.
Choosing that the $x$-direction is normal to the \gls{dw} plane for the $(100)_\text{pt}$ walls, the $\phi_y\psi_z - \phi_x$ wall (reversing $\phi_y$ and  $\psi_z$ in one layer and $\phi_x$ in the other) leads to a mirror symmetry across the boundary, while a $\phi_x\psi_z - \phi_y$ wall exhibits a \qty{90}{\degree} rotation. 
This excludes the $(001)_\text{pt}$ plane where $\phi_x$ and $\phi_y$ are indistinguishable.
Overall, this establishes the existence of four possible \glspl{dw} which should be representative for the entire configurational space (\autoref{tab:tilts}).

\begin{table}
\centering
\caption{All symmetry-distinct \glspl{dw} associated with any wall plane.
Every wall is named according to which tilt phases change sign at the boundary in each respective layer, and whether they are \qty{180}{\degree} (ferroelectric) or \qty{90}{\degree} (ferroelectric and ferroelastic).
The paired columns refer to the two layers.}
\label{tab:tilts}
\begin{threeparttable}
\begin{tabular}{l*{9}c}
\toprule
Wall type  & \multicolumn{2}{c}{$Q_{\phi_x}$ }&& \multicolumn{2}{c}{$Q_{\phi_y}$} && \multicolumn{2}{c}{$Q_{\psi_z}$} & broken tilt phases \\
\midrule\addlinespace[1ex]
$\mathbf{180^\circ}$ \textbf{\gls{dw}} \\\addlinespace[1ex]
   \hspace{2mm}$\phi_x\phi_y$  & $-1$ & $-1$ & & $-1$ &  $-1$ & &\hspace{7pt}1 &\hspace{7pt}1 & 2\hspace{5mm}2\\\addlinespace[1ex]
   \hspace{2mm}$\psi_{z}$ & \hspace{7pt}1 & \hspace{7pt}1 && \hspace{7pt}1 & \hspace{7pt}1 && $-1$ & $-1$ & 1\hspace{5mm}1\\\addlinespace[1ex]
$\mathbf{90^\circ}$ \textbf{\gls{dw}} \\\addlinespace[1ex]
    \hspace{2mm}$\phi_x\psi_z-\phi_y$    & $-1$ & \hspace{7pt}1 &&  \hspace{7pt}1 & $-1$ && $-1$ & \hspace{7pt}1& 2\hspace{5mm}1\\\addlinespace[1ex] 
     \hspace{2mm}$\phi_y\psi_z-\phi_x$    &  $-1$ & \hspace{7pt}1 && \hspace{7pt}1 & $-1$ && \hspace{7pt}1 & $-1$ & 1\hspace{5mm}2\\\bottomrule
\end{tabular}
\end{threeparttable}
\end{table}

\subsection*{Domain wall energetics and profiles}

For this study, the \glspl{dw} listed in \autoref{tab:tilts} were created with four $(100)_\text{pt}$, four $(110)_\text{pt}$ and three distinct $(001)_\text{pt}$ planes, amounting to a total of 11 different \glspl{dw} (in the $(001)_\text{pt}$ planes, the \qty{90}{\degree} \glspl{dw} become symmetrically equivalent).
Depending on the geometry of the wall, there is a bound surface charge density at the interface, $\sigma_b = (P_2 - P_1)\cdot \vec{n}_{\perp \text{DW}}$, if there exists a discontinuous polarization component normal to the \gls{dw} plane.
This is usually described as the walls having a head-to-head (H--H) or tail-to-tail (T--T) configuration \cite{bednyakov2018physics}.
In \ce{Ca3Ti2O7}, three categories can be used to describe the various charge states, illustrated in \autoref{fig:fig2}b. 
If the polarization is parallel to the \gls{dw}, there is no bound charge and the wall is electrostatically neutral.
Meanwhile, the charged walls can be distinguished by whether the polarization is at a \qty{45}{\degree} or \qty{90}{\degree} to the wall plane.
These two cases will therefore be referred to as partially charged and charged to distinguish them relative to one another.
Charged \glspl{dw} of any kind are highly unstable due to unfavorable electrostatic fields and are therefore rare except in improper ferroelectrics where \gls{dw} formation is geometrically driven, such as the hexagonal manganites \cite{EvansGarciaMeierBibes+2020, schaab2014imaging} or \gls{rp} phases.

The domain structure in undoped \ce{Ca3Ti2O7} has been experimentally observed to consist of parallel \qty{90}{\degree} \glspl{dw}, interspersed with \qty{180}{\degree} \glspl{dw} inside each domain.
The planar distribution shows that they predominantly exist within the $\{100\}_\text{pt}$ planes, while A-site doping with \ce{Sr} stabilizes additional walls in the $\{110\}_\text{pt}$ plane \cite{nakajima2019charged, oh2015experimental}.
These only exist as \qty{180}{\degree} ferroelectric \glspl{dw} due to the lattice mismatch in the orthorhombic structure, and are either charged or neutral.
\Glspl{dw} in the $(001)_\text{pt}$ planes, parallel to the perovskite layers, have also been observed in several \gls{rp} systems like \ce{(Ca,Sr)3Ti2O7} and \ce{Ca3Ru2O7} \cite{lei2018observation, nakajima2021charged}. Examples of \gls{dw} structures for each plane are shown in \autoref{fig:fig2}c--e.

Three \glspl{mlip} were created based on a training set with \gls{dft} evaluated structures, as described in the Methods section.
One was trained only on structural information through angular and radial descriptors, while the other two use the \gls{qnep} framework, incorporating electrostatic contributions to the total energy \cite{song2024general, fan2026qnep}.
The two \gls{qnep} models differ by the inclusion of an additional \gls{dft} dataset with target \glspl{bec}. 
All three models were used to relax each of the structures and evaluate formation energies, following the exact same procedure as the \gls{dft} calculations. 
Although there are only four symmetry-distinct \gls{dw} configurations, any gradient-dependent properties, such as the local strain gradient, order parameter, or polarization profiles depend on the orientation of the wall. 
All the structures are therefore different both in terms of electrostatics and local wall geometry.

\begin{table*}
\centering
\caption{
Formation energies of the 11 different \glspl{dw} considered in this work, calculated with \gls{dft} and the three \glspl{mlip}.
The best agreement with \gls{dft} is found for the \gls{qnep} model trained with target \glspl{bec}.
$P_{\perp \text{DW}}$ indicates whether the normal components of the polarization vectors meeting at the wall point toward or away from each other (H--H/T--T) or are continuous (H--T).
The wall width ($w_\text{DW}$) is determined from order parameter profiles.
See \autoref{tab:tilts} for how each wall type is categorized in terms of tilt phases.
}
\label{tab:bigtable}
\begin{threeparttable}
\begin{tabular}{llccccccc}
\toprule
Plane & Wall type & \multicolumn{4}{c}{$E_{\text{DW}}^f$ (\unit{\milli\joule\per\meter\squared})\tnote{*}} & $P_{\perp\text{DW}}$ & $w_\text{DW}$ & Cell length\tnote{\dag}  \\ \addlinespace[1ex]
  & & DFT & NEP & qNEP & qNEP with BECs & & (\unit{\angstrom}) & 
  (u.c.)\\
\midrule\addlinespace[1ex]
$(100)_\text{pt}$ & \qty{180}{\degree} \gls{dw}: $\phi_x\phi_y$ & 187 & 160 & 190 & 186 & H--H/T--T & 18 & 6  \\
& \phantom{\qty{180}{\degree} DW:} $\psi_z$ & 155 & 146 & 167 & 158 & H--H/T--T & 19 & 6 \\
& \:\:\qty{90}{\degree} \gls{dw}: $\phi_x\psi_z-\phi_y$ & 122 & 122 & 138 & 126 & H--T & 15 & 6\\
& \phantom{\qty{180}{\degree} DW:} $\phi_y\psi_z-\phi_x$ & 142 & 128 & 146 & 135 & H--H/T--T & 20 & 6 \\ \addlinespace[1ex]
$(110)_\text{pt}$ & \qty{180}{\degree} \gls{dw}: $\phi_x\phi_y$ & 208 & 181 & 206 & 204 & H--H/T--T & 20 & 8\\
& \phantom{\qty{180}{\degree} DW:} $\psi_z$ & 188 & 159 & 185 & 174 & H--H/T--T & 20 & 8 \\ 
 & \phantom{\qty{180}{\degree} DW:} $\phi_x\phi_y$ & 199 & 185 & 195 & 197 & (neutral) & 19 & 8\\
& \phantom{\qty{180}{\degree} DW:} $\psi_z$ & 137 & 136 & 143 & 136 & (neutral) & 14 & 8 \\\addlinespace[1ex]
$(001)_\text{pt}$ & \qty{180}{\degree} \gls{dw}: $\phi_x\phi_y$ & 145 & 110 & 134 & 151 & (neutral) & 13.5 & 4\\
& \phantom{\qty{180}{\degree} DW:} $\psi_z$ & 13 & 2.5 & 16 & 13 & (neutral) & 6 & 4\\
& \:\:\qty{90}{\degree} \gls{dw}: $\phi\psi_z-\phi$ & 82 & 69 & 90 & 87 & (neutral) & 13.5 & 4 \\
\bottomrule
\end{tabular}
\begin{tablenotes}\footnotesize
\item[*]
    For each wall all methods use the same supercell, so the comparison between \gls{dft} and the \glspl{mlip} is consistent.
    Note that the absolute formation energies of the charged and partially charged \glspl{dw} are inherently cell-size-dependent \cite{SmabratenCharged2018}.
\item[\dag]
    Supercell length in terms of the number of unit cells normal to the \gls{dw} plane.
    A 48-atom unit cell was used for the $(110)_\text{pt}$ \glspl{dw}, and a 96-atom cell was used for $(100)_\text{pt}$ and $(001)_\text{pt}$ \glspl{dw}. 
\end{tablenotes}
\end{threeparttable}
\end{table*}

In general, the formation energies (\autoref{tab:bigtable}) show an increase in cost with the number of broken tilt phases and the larger structural distortion this entails, as well as with the wall having an H--H or T--T component of the polarization.
The ferroelectric $\phi_x \phi_y$ wall, which switches the phase of both $\phi_x$ and $\phi_y$ in both layers, consistently has the largest formation energy within each plane.
Meanwhile, for the completely neutral $(001)_\text{pt}$ plane, the $\psi_z$ \gls{dw} has an almost negligible energy cost compared to the other walls.
The formation energies mostly range from about \num{100} to \qty{200}{\milli\joule\per\meter\squared}, which is a typical magnitude for many ferroic systems, such as the \qty{180}{\degree} \glspl{dw} in \ce{PbTiO3} (\qtyrange{120}{160}{\milli\joule\per\meter\squared}) \cite{ChandrasekaranDefect2013, zhang2023first} or structural antiphase boundaries in \ce{PbZrO3} (\qty{190}{\milli\joule\per\meter\squared}) \cite{wei2014ferroelectric}.
For the $(100)_\text{pt}$ ferroelastic \glspl{dw}, the formation energy is close to the average of perovskite \ce{CaMnO3}, isostructural to\ce{CaTiO3}, where \glspl{dw} with one or two broken tilt phases have calculated formation energies of \qty{67}{\milli\joule\per\meter\squared} and around \qty{163}{\milli\joule\per\meter\squared} \cite{skogvoll2025local}, respectively.

\begin{figure}
    \centering
    \includegraphics[width=\linewidth]{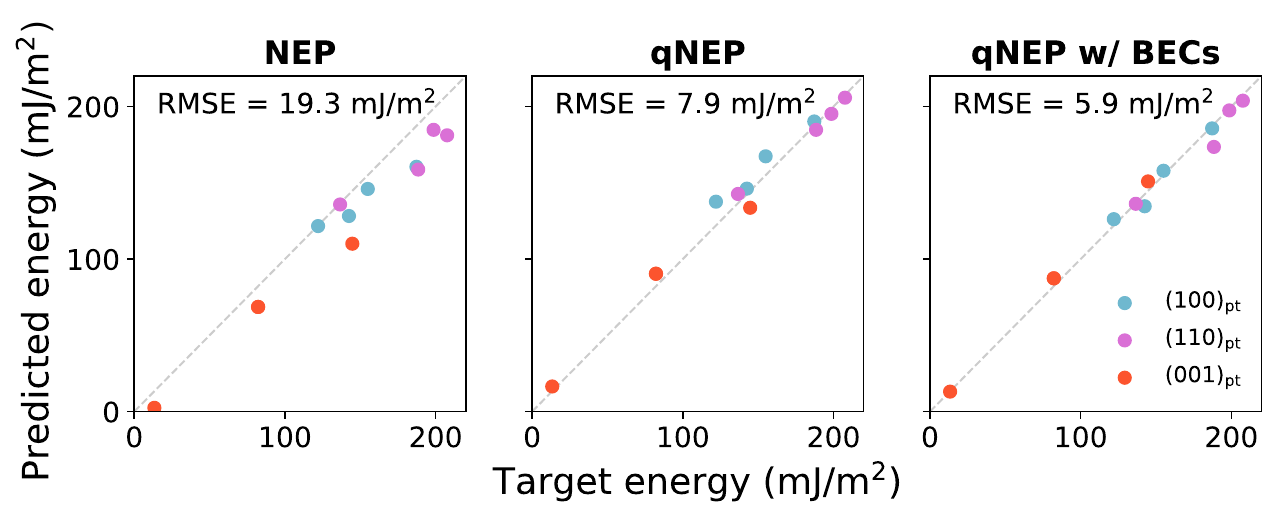}
    \caption{
    Parity plots comparing the predicted formation energies of \gls{dw} structures to target energies from \gls{dft}.
    The calculated \acrfull{rmse} is lowered from \qty{19.3}{\milli\joule\per\meter\squared} to  \qty{5.9}{\milli\joule\per\meter\squared}, amounting to a \qty{70}{\percent} decrease when training using the \gls{qnep} framework with a dataset including target \acrfullpl{bec}.
    }
    \label{fig:formation_energies}
\end{figure}

The comparison between target and predicted formation energies for all three models is presented in \autoref{fig:formation_energies}.
With the conventional \gls{nep}, the formation energies are systematically underestimated even though the relaxed walls are structurally almost identical to those obtained with \gls{dft}.
The closest agreement is for the $\phi_x\psi_z-\phi_y$ wall in the $(100)_\text{pt}$ plane, which is the only wall in the set with a continuous normal component of the polarization (H--T) and which \gls{nep} reproduces exactly.
This is because the total energy is only calculated as a sum of local site energies.
When using \gls{qnep}, there is a considerable improvement in the formation energies relative to \gls{dft} values. 
Using the same dataset, trained with the \gls{qnep} architecture, the \gls{rmse} is lowered from \qty{19.3}{\milli\joule\per\meter\squared} to \qty{7.9}{\milli\joule\per\meter\squared}.
An extended dataset with target \glspl{bec} lowers the \gls{rmse} further to \qty{5.9}{\milli\joule\per\meter\squared}, a \qty{70}{\percent} reduction relative to the conventional \gls{nep} model. Meanwhile, the overall \gls{rmse} in target vs. predicted energies for the entire dataset is not lowered for any of the \gls{qnep} potentials compared to regular \gls{nep} (for the full parity plots of each model see Figs.~\ref{fig:SM_parity_model1}--\ref{fig:SM_parity_model3} in the Supplementary Information \cite{SM}).
This demonstrates the benefit of a charge-aware treatment for the interfaces considered here, particularly when long-range electrostatics are important. 
 
\begin{figure*}
    \centering
    \includegraphics[width=\linewidth]{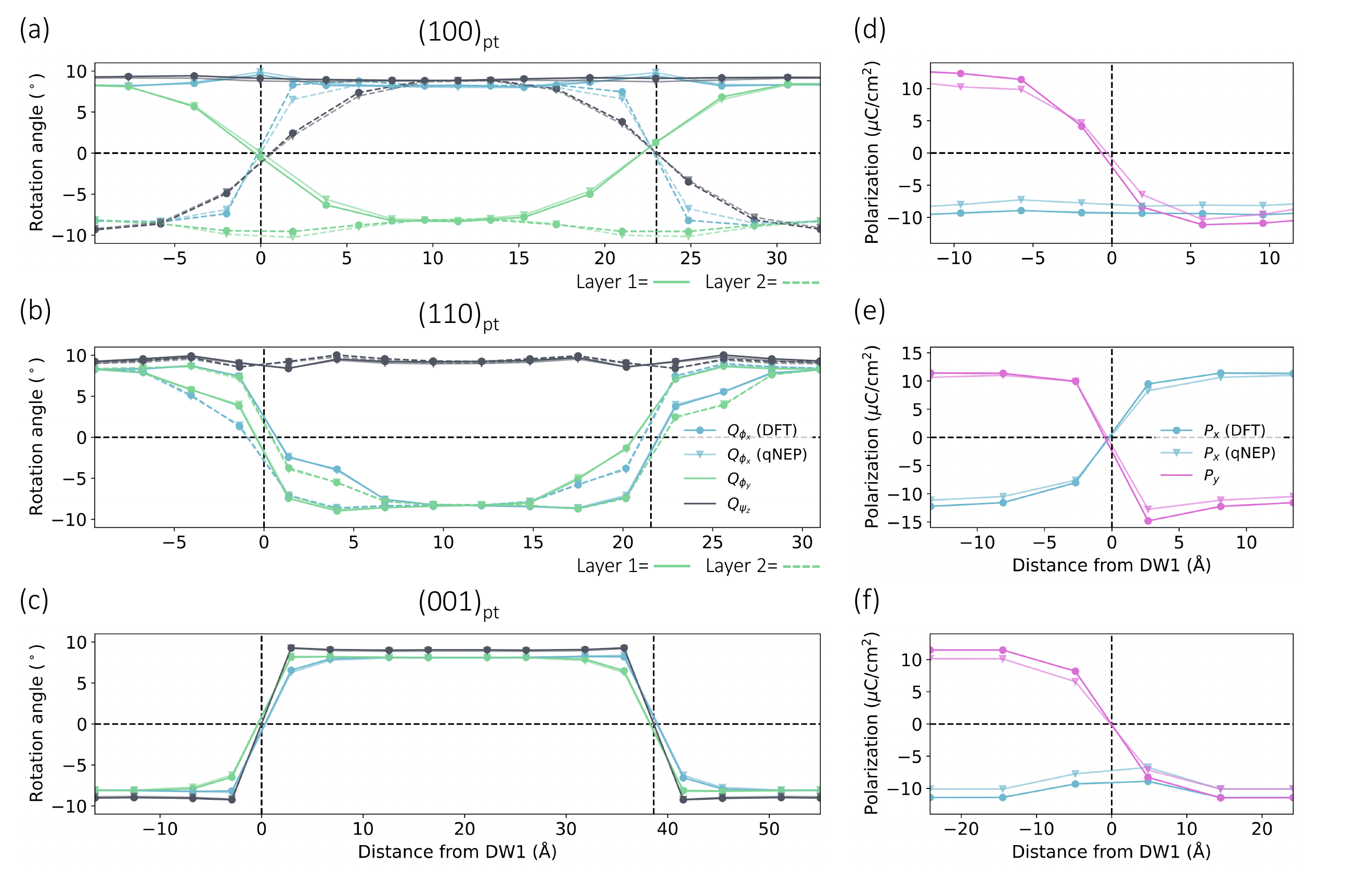}
    \caption{
    Order parameter profiles for the (a) $\phi_x\psi_z-\phi_y$ \gls{dw}, (b) the $\phi_x\phi_y$ neutral wall and (c) the $\phi\psi_z-\phi$ wall.
    The mode amplitude ($Q$) is calculated as the rotation angle around the $x$, $y$ and $z$ axes, except for the $(110)_\text{pt}$ wall where the tilt axis is at $45^\circ$ with the lattice axes. 
    (d--f) Polarization profiles corresponding to the first \gls{dw} (DW1) in the supercells used for order parameter plots. 
    Note that for the $(001)_\text{pt}$ wall it is not possible to meaningfully assign a phase to the tilts due to the asymmetry between the perovskite layers. 
    The positions of \glspl{dw} are indicated with vertical dashed lines.
    Throughout, \gls{dft} results are shown in the darker and \gls{qnep} results in the lighter shade of each color.
    }
    \label{fig:profiles}
\end{figure*}

The \gls{dw} structures relaxed with \gls{dft} and the final \gls{qnep} model were analyzed by calculating the order parameter and polarization profiles.
The order parameter profile shows the evolution of the $X_2^+$ and $X_3^-$ \glspl{irrep}, and is calculated as the rotation angle around each of the axes.
\autoref{fig:profiles}a--c shows the profiles for one neutral \gls{dw} structure in each plane, while the corresponding polarization profiles are displayed in \autoref{fig:profiles}d--f. 
Depending on whether the tilting changes phase or not, the order parameter follows a $\tanh$ or $\sech$ profile across the wall, consistent with Landau theory \cite{salje1990phase}.
The $(001)_\text{pt}$ walls have significantly less distortion and are notably narrower than those in the other planes, reflecting the weaker interaction between perovskite layers.
Across all six panels of \autoref{fig:profiles} the \gls{qnep} and \gls{dft} order parameter profiles are very close, reproducing the wall position, the wall width and the approach to the bulk plateau; the only visible deviation is a slight broadening of $Q_{\phi_x}$ at the $(110)_\text{pt}$ wall.
The polarization profiles likewise agree in shape and wall width, and differ only by a nearly uniform offset of about \qty{1}{\micro\coulomb\per\centi\meter\squared} in magnitude, corresponding to roughly \qty{10}{\percent} of the bulk value.
This offset originates in the predicted \glspl{bec} of the reference $I4/mmm$ structure, which are systematically lower than those from \gls{dft} (\autoref{fig:SM_parity_BEC}), rather than in the wall structure itself.
Recomputing the polarization with the actual \glspl{bec} of the relaxed supercells from \gls{qnep} gives the same profiles, with the magnitude reduced by about \qty{8}{\percent}.

We remark that for the charged and partially charged \glspl{dw}, the H--H and T--T components of the polarization induce a highly unfavorable internal electric field. 
As a result there is an attractive force between the walls, and the formation energy does not converge with cell size.
This is also found in other computational studies of charged \glspl{dw}, where the gradient of the electrostatic potential is independent of \gls{dw} distance \cite{SmabratenCharged2018}. 
With a conventional \gls{nep} this dependence is absent altogether once the walls are separated by more than the descriptor cutoff, which is in itself a signature of the long-range character of the interaction.
For the H--H/T--T walls in this study, the electrostatic potential was lowered by the \ce{Ti} atoms reversing their off-centering from the \ce{TiO6} center of mass, even if parts of the mid-domain were frozen during relaxation with both \gls{dft} and \gls{qnep}. 
It is therefore expected that the true formation energies for these walls are even higher. 

\subsection*{Atomistic switching mechanisms}
Because of the hybrid improper mechanism for switching in \ce{Ca3Ti2O7}, both \qty{90}{\degree} and \qty{180}{\degree} \gls{dw} movement requires entire octahedra to change their rotational patterns, and is thus associated with high coercive fields for switching. 
Despite this, a range of field strengths have been reported, mostly above \qty{100}{\kilo\volt\per\centi\meter}, though also as low as \qty{5}{\kilo\volt\per\centi\meter} \cite{zhang2021review, li2017ultra}. 
There are many factors influencing the field magnitude, but most fundamental is the atomistic switching pathway with which the \gls{dw} moves. 
Knowing this intrinsic mechanism is also essential to design effective approaches to lowering coercive fields, while retaining a stable polarization. 

Several studies have enumerated these minimum switching pathways and calculated the associated energy barriers from bulk structures \cite{nowadnick2016domains, Munro2018Discovering, li2020suppressing}. However, this process requires prior knowledge of the full order parameter space and it is difficult to predict how these switching mechanisms unfold in an actual \gls{dw}. 
The trained \gls{qnep} potential is a convenient alternative to determine these minimum energy pathways because any \gls{dw} propagation transpires through the lowest-energy configuration, with nearly the precision of \gls{dft}. 
One can then also investigate the dynamics of wall motion, such as in- or out-of-plane correlation or nucleation events. 
 
To determine exactly how each of the four \glspl{dw} preferentially moves, \gls{md} simulations were conducted on long supercells with 48 \ce{CaTiO3} units in the $x$-direction, perpendicular to the $(100)_\text{pt}$ wall plane.
Wall motion can then be ascertained from thermal fluctuations around the positional average.
The \gls{dw} is topologically protected against spontaneously vanishing, and can only annihilate in the case that it meets the wall on the opposite side of the cell.
While the mobility depends on temperature and system size, the switching pathway does not as it remains the same at every temperature and cell size examined.
The mobility decreases as the supercell is enlarged in the wall plane, approaching a small value in the large-cell limit (\autoref{fig:SM_DW_pos}).
This reflects the fact that the walls are not rigid: they roughen within their own plane, so that, depending on the wall type, different segments fluctuate largely independently and the displacement of the wall as a whole is an average that is increasingly suppressed as the wall grows.
In the presence of an applied field the mobility will instead be dominated by the field-driven motion.
Wall trajectories from these simulations, including a case in which two walls annihilate, are shown in \autoref{fig:SM_motion}.

\begin{figure}
    \centering
    \includegraphics[width=\linewidth]{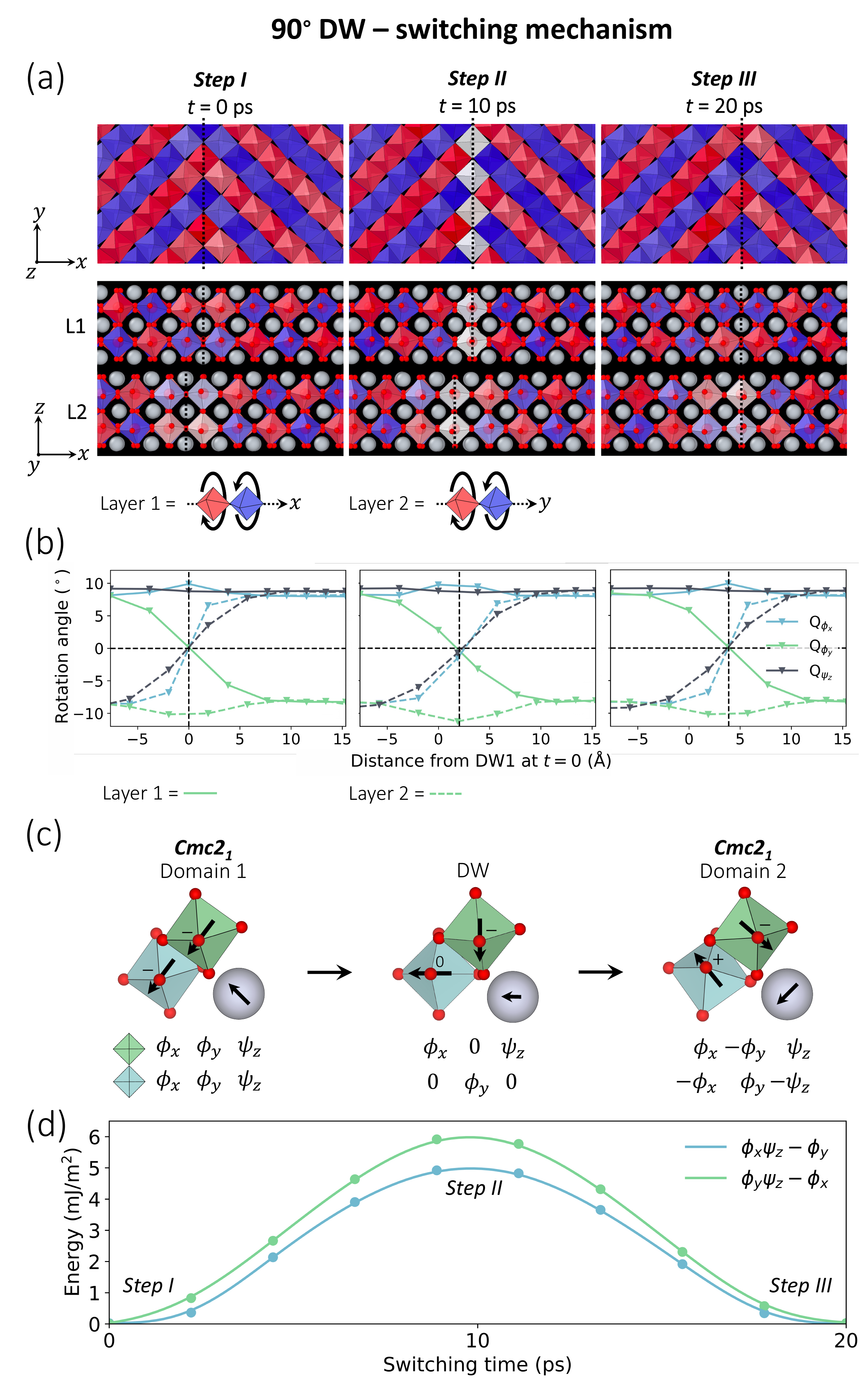}
    \caption{
    (a) \gls{md} simulations of the displacement of a $\phi_x\psi_z-\phi_y$ wall at \qty{300}{\kelvin}.
    In the top panels, every octahedron is color-coded with respect to the $\phi_x$ rotation, clearly showing the rotation by \qty{90}{\degree} at the wall.
    For the bottom panels, layer 1 (L1) and layer 2 (L2) use the rotations around $x$ and $y$ to accentuate the position of the wall.
    (b) The order parameter profiles corresponding to step I, II and III in (a), with the solid lines representing layer 1 and dashed lines representing layer 2.
    (c) Symmetries of each domain state and the instantaneous wall symmetry, illustrated with the change in tilting and polarization across the boundary.
    (d) \Gls{neb} calculations of the energy barrier for switching each of the two \qty{90}{\degree} wall types.
    }
    \label{fig:FA}
\end{figure}

\paragraph*{$\mathbf{90^\circ}$ \textbf{\glspl{dw}}.}
For the two different \qty{90}{\degree} walls, $\phi_x\psi_z-\phi_y$ and $\phi_y\psi_z-\phi_x$, the switching was observed to proceed through a one-step mechanism where individual octahedra reverse their tilting. 
Due to the asymmetry in these \glspl{dw} (see \autoref{tab:tilts}), this involves reversing two tilting phases in one layer, and a single phase in the other. 
Despite this, the displacement remains coherent across the stacking direction, even with 20 layers stacked along $z$. 
The simulation of a $\phi_x\psi_z-\phi_y$ wall displacing a single step at \qty{300}{\kelvin} can be seen in \autoref{fig:FA}a, with the evolution in tilting angles plotted in \autoref{fig:FA}b.
The intermediate configuration throughout the process is $ \phi_x 0 \psi_z \: 0\phi_y0$, which in bulk would be consistent with a polar $Amm2$ symmetry, as illustrated in \autoref{fig:FA}c. 
For the $\phi_y\psi_z-\phi_x$ \gls{dw} this switching procedure looks exactly the same, only with $\phi_x$ and $\phi_y$ in opposite layers. 

It is important to note that although the instantaneous wall symmetry can be characterized in terms of a three-dimensional space group, a \gls{dw} is a quasi-two-dimensional object and therefore inherently symmetry-broken. 
The energy difference between the bulk ($Cmc2_1$) and wall symmetry therefore does not necessarily relate to the energy cost of moving a pre-existing wall.
The energy cost for switching was estimated using the \gls{neb} method with the \gls{qnep} potential (\autoref{fig:FA}d), resulting in a barrier of \qty{4.9}{\milli\joule\per\meter\squared} for the $\phi_x\psi_z-\phi_y$ wall and \qty{5.9}{\milli\joule\per\meter\squared} for the $\phi_y\psi_z-\phi_x$ wall, which is considerably lower than the formation energies. 
This minimum energy switching mechanism is qualitatively consistent with the orthorhombic twin switching in Ref.~\citenum{nowadnick2016domains}, but it is not possible to directly compare the barrier energies.

\begin{figure*}
    \centering
    \includegraphics[width=\linewidth]{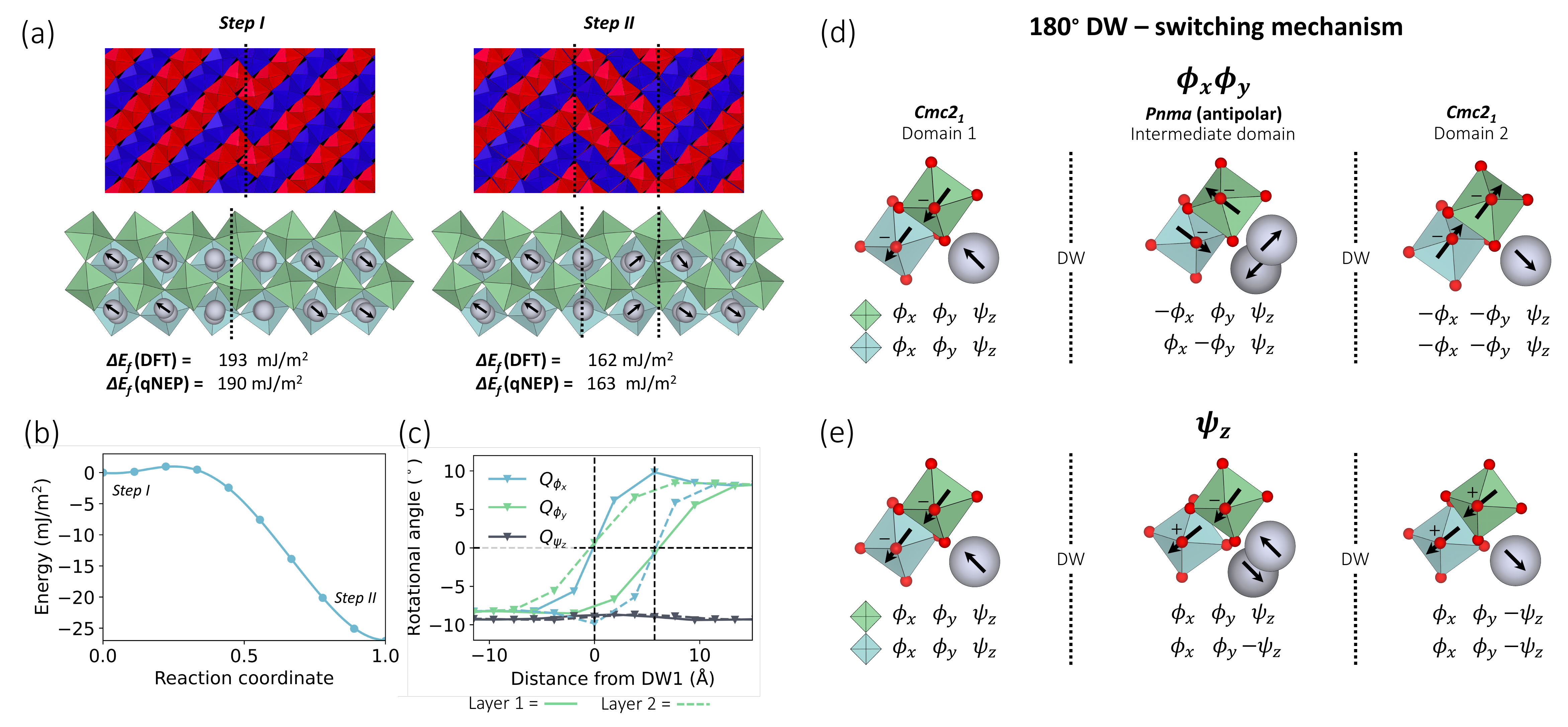}
    \caption{(a) The $\phi_x\phi_y$ \gls{dw} before (step I) and after (step II) the splitting process, along with the formation energies for each configuration calculated with \gls{dft} and \gls{qnep}. 
    (b) \gls{neb} calculations of an approximate energy barrier that separates a metastable non-split (step I) and a stable split configuration (step II). This is not an energy barrier resulting from switching or \gls{dw} motion. 
    (c) The order parameter profile depicting the asymmetry between each layer, producing an antipolar $Pnma$ symmetry in between the two coupled ferroelastic walls. 
    The atomistic switching mechanism for the (d) $\phi_x\phi_y$ and (e) $\psi_z$ ferroelectric walls. 
    They both switch through a two-step process, but for $\psi_z$ the polarization flips by \qty{180}{\degree} in one perovskite layer at a time. 
    Two arrows are used when there is a difference in polarization direction between the top (lighter gray sphere) and bottom layers (darker gray sphere).
    }
    \label{fig:FE}
\end{figure*}

\paragraph*{$\mathbf{180^\circ}$ \textbf{\glspl{dw}}.}
On heating from their \qty{0}{\kelvin} relaxed structures, the $\phi_x \phi_y$ and $\psi_z$ ferroelectric \glspl{dw} turn out to be trapped in metastable local minima when the analysis starts from a sharp wall configuration. 
The true ground state for the $\phi_x \phi_y$ \gls{dw} is a configuration where the \qty{180}{\degree} wall is split up into two low-energy ferroelastic-type boundaries connected by symmetry. 
This means that at each split boundary, only a single out-of-phase rotational mode reverses in each layer, producing an antipolar mid-domain and effectively lowering the electrostatic energy. 
The size of this splitting was found to increase with temperature, growing from about \qty{25}{\angstrom} at \qty{0}{\kelvin} to about \qty{30}{\angstrom} at \qty{150}{\kelvin}. 
The simulated splitting process is depicted in \autoref{fig:FE}a, with the energies of the relaxed split and non-split structures calculated with \gls{dft} and \gls{qnep}. 
\numproduct{8x1x1} supercells were used to accommodate the widening of the \glspl{dw}, which explains the discrepancy between these formation energies and those of \autoref{tab:bigtable}. 

When a \qty{180}{\degree} \gls{dw} splits in this way, it creates two interfaces that can move somewhat independently both of each other and between layers. 
The total switching barrier can then be lowered by sequentially moving one element of the wall at a time, as opposed to moving the whole wall coherently. 
This could explain the low coercive field observed for ferroelectric switching. Unfortunately, it was not feasible to determine a definitive splitting barrier using \gls{neb} because there are multiple moving boundaries in a supercell with two split \gls{dw} configurations.
However, the qualitative transition from one local minimum to a global minimum is plotted in \autoref{fig:FE}b. 
This does not show the energy cost from \gls{dw} movement, but instead demonstrates the inherent limitations that come with manually creating and relaxing these structures with \gls{dft}.
We note that this splitting was present also for the regular \gls{nep} model without charges.
The calculated evolution of the order parameters is presented in \autoref{fig:FE}c.

The atomistic switching mechanism is a transition from $Cmc2_1$ to an antipolar $Pnma$ domain where the polarization points in opposite directions in adjacent perovskite slabs, and where the instantaneous symmetry of the \qty{90}{\degree} boundaries corresponds to $Pmn2_1$ in bulk, see \autoref{fig:FE}d.
The size of the splitting is essentially a competition between the cost of the boundaries and the volume of the $Pnma$ domain.
This is analogous to extended dislocations in face-centered cubic metals, where two partial dislocations surrounding a stacking fault reach an equilibrium distance, balancing repulsive and attractive forces \cite{hull2011introduction}.
Within ferroelectrics, such splitting is an instance of the domain wall decomposition reactions recently proposed as a general feature of multiple-order-parameter systems \cite{Zhou2025domainwall}.

The $\psi_z$ ferroelectric \gls{dw} exhibits the same kind of splitting, where the points of reversal of in-phase rotations in each layer move away from each other.
This creates a \qty{180}{\degree} switching of the polarization in one perovskite slab at a time, resulting again in a $Pnma$ mid-domain. These boundaries are not associated with any strain as for the previous $\phi_x \phi_y$ \gls{dw}.
The instantaneous \gls{dw} symmetry is $Pmc2_1$. 
When $(100)_\text{pt}$  $\psi_z$ \glspl{dw} inside consecutive layers move away from one another, this creates a $(001)_\text{pt}$  $\psi_z$ \gls{dw} in the rocksalt layer between them. 
Since these have a very low formation energy (see \autoref{tab:bigtable}), these walls exhibit a much higher mobility and have less inter-layer coherence than $\phi_x\phi_y$ walls. Order parameter profiles for the non-split $\phi_x\phi_y$ configuration and the two \qty{180}{\degree} $(001)_\text{pt}$ \glspl{dw} can be found in Figs.~\ref{fig:SM_OP_unsplit} and \ref{fig:SM_OP_001} in the Supplementary Information \cite{SM}. 

The identified atomistic switching mechanisms are consistent with the lowest energy switching pathways found in Ref.~\citenum{nowadnick2016domains}, and offer a two-step pathway through an antipolar phase as a mechanism for the unexpectedly low coercive fields reported for \ce{Ca3Ti2O7}. 
Other switching pathways through alternative rotation patterns have been proposed that are not observed here \cite{Munro2018Discovering, li2020suppressing}. 
\gls{md} simulations of charged and neutral ferroelectric walls in the $(110)_\text{pt}$ planes at finite temperature show the exact same splitting (\autoref{fig:SM_110_split}), although a full analysis of these walls was not considered in the current work. 
Antipolar intermediate domains have also been experimentally observed to screen head-to-head and tail-to-tail \glspl{dw} in \ce{(Ca,Sr)3Ti2O7} \cite{Lee2017}.

\section*{Discussion}
We have demonstrated that \glspl{mlip} based on the \gls{nep} framework can accurately reproduce the local geometry and electrostatics of \glspl{dw} in \ce{Ca3Ti2O7}. 
The charge-aware \gls{qnep} models also systematically lower the error in the \gls{dw} formation energies.
The order parameter profiles are almost identical to those from \gls{dft}, and the polarization profiles are reproduced up to a small offset in magnitude. 
\Gls{md} simulations establish the switching mechanisms of the four symmetry-distinct $(100)_\text{pt}$ \glspl{dw} in \ce{Ca3Ti2O7}, revealing low-energy pathways for both inter-layer and intra-layer \gls{dw} movement.
The \qty{90}{\degree} \glspl{dw} display a polar intermediate structure, while the \qty{180}{\degree} \glspl{dw} split into two boundaries enclosing an antipolar domain and switch through a two-step mechanism. 
These atomistic simulations thereby establish a microscopic mechanism of domain wall motion. 

Beyond accuracy, \gls{qnep} offers a computational advantage for exploring \glspl{dw} in layered perovskite systems with large unit cells, at a fraction of the computational cost of \gls{dft}. 
Even the smaller supercells of the \gls{rp} phase are at the limit of what is feasible with \gls{dft}, while \gls{qnep} can obtain optimized wall geometries for significantly bigger cells within minutes.
This is ideal for systematically exploring stabilization mechanisms of charged \glspl{dw} in \ce{(Ca,Sr)3Ti2O7} or other compositionally tuned \gls{rp} phases.
The fast optimization could also be utilized for additional phenomena such as magnetoelectric coupling, for example by using isolated \gls{dw} structures as input for further analysis with \gls{dft}. 
Furthermore, the efficiency of the \gls{nep} method paves the way for field-driven large-scale simulations of nucleation, growth and switching dynamics.

\section*{Methods}

\subsection*{Density functional theory}

\Gls{dft} calculations were used both to evaluate the training data and to determine the formation energies and local geometries of the various \glspl{dw}.
They were performed using the \textsc{vasp} \cite{Kresse1996, Kresse1999} code and the PBEsol functional (Perdew--Burke--Ernzerhof revised for solids), within the generalized gradient approximation \cite{Perdew2008}.
The electronic wave functions are represented using the projector augmented wave approximation \cite{Blochl1994}, where \ce{Ca}$(3s^2, 3p^6, 4s^2)$, \ce{Ti}$(3p^6, 3d^3, 4s^1)$ and \ce{O}$(2s^2, 2p^4)$ were treated as valence electrons.
The plane-wave cutoff energy was set to \qty{550}{\electronvolt}.

A $\varGamma$-centered \numproduct{9x9x9} $k$-point mesh was used for the primitive cubic cell of \ce{CaTiO3}, with an equivalent $k$-point density for all other cell sizes.
For bulk structures, atomic positions were relaxed until the residual forces on all atoms were below  \qty{1}{\milli\electronvolt\per\angstrom}.
For the \gls{dw} structures, which contain between \num{192} and \num{768} atoms depending on the wall type, the residual forces were relaxed below \qty{20}{\milli\electronvolt\per\angstrom}.
The same threshold was applied in the corresponding \gls{mlip} relaxations, so that the comparison between methods is not affected by the choice of tolerance.
The supercells were relaxed with fixed lattice parameters such that they remain equal to the bulk \gls{rp} phase.

\Gls{dw} formation energies were calculated following
\begin{align}
    E_{\text{DW}}^f = \frac{1}{2A}(E_{\text{DW}} - E_{\text{bulk}}),
\end{align}
where $E_{\text{DW}}$ is the total energy of a supercell with two \glspl{dw}, and $E_{\text{bulk}}$ is the total energy of a corresponding monodomain supercell.
$A$ is the cross-sectional area of the wall.
\glspl{bec} were calculated using density functional perturbation theory.

For the creation and analysis of the \gls{dw} structures, the exact same procedure was also used with the \glspl{mlip}.
Polarization profiles were calculated using \glspl{bec} from the high-symmetry $I4/mmm$ structure, with a mid-layer \ce{Ca} atom as the unit cell origin.
Phonon modes were analyzed with the finite difference method implemented in \textsc{phonopy} \cite{togo2023first, togo2023implementation}.
Mode decomposition of the ground state structure was calculated with \textsc{isodistort} \cite{isotropy, campbell2006isodisplace}.
All crystal structures are visualized using \textsc{vesta} \cite{Momma:db5098}.
Optimized supercell structures can be retrieved on Zenodo \cite{zenodo}.

\subsection*{Neuroevolution potential model}
An \gls{mlip} was constructed using the \gls{nep} scheme \cite{song2024general} implemented in \textsc{gpumd} \cite{fan2022gpumd, XuBuPan25}, following the procedure outlined in Ref.~\citenum{fransson2023phase}.
The model was trained iteratively, starting with a selection of ideal, strained, and rattled perovskite and \gls{rp} structures with one to four octahedral layers per perovskite slab, adding snapshots from \gls{md} runs in the second iteration.
For the rattled structures, atoms were randomly displaced using a Monte Carlo procedure supplied by the \textsc{hiphive} package \cite{eriksson2019hiphive}.
In addition, ferroelastic \gls{dw} structures in \ce{CaTiO3} and a subset of \glspl{dw} in \ce{Ca3Ti2O7} were included.
Every structure was evaluated in terms of energies, forces, and stresses with \gls{dft}. 

Three different models were trained, one using the regular \gls{nep} framework based on radial and angular descriptors, and two \gls{qnep} models using a charge-aware approach treating partial charges as latent and learnable model features \cite{fan2026qnep}.
The difference between the latter two models is the inclusion of an additional dataset with evaluated \glspl{bec}, bringing the training set of the final potential to \num{760} structures.
The \textsc{ase} \cite{larsen2017atomic} and \textsc{calorine} \cite{lindgren2024calorine} packages were used to prepare the training data and for post-processing.
Energy barriers for switching pathways were calculated using the \gls{neb} method \cite{henkelman2000climbing} as implemented in \textsc{ase} \cite{kolsbjerg2016automated}.
Ten images were relaxed at fixed volume using the final \gls{qnep} model and the BFGS optimizer. 
To avoid any additional contributions from wall--wall interactions to the energy profile, a stiff spring constant of \qty{10}{\electronvolt\per\angstrom\squared} was employed, ensuring that both walls moved coherently. 
In all cases, the lowest-energy path that still corresponded to the isolated movement of a single wall was used.
For additional information on the models, the training set and validation, see the Supplementary Information \cite{SM, Biztek2006Structural, guenole2020assessment, ALI20052867, Carpenter2007, Gu2012, Zhang2025, yoshida2018hybrid, gradauskaite2026revival}.

\subsection*{Molecular dynamics}
\Gls{md} simulations were carried out with the \textsc{gpumd} code.
The NPT ensemble was consistently used in all simulations, with temperature and pressure controlled through the velocity-rescaling and stochastic cell-rescaling methods \cite{Bussi2007Canonical, Bernetti2020Pressure}.
To evaluate the phase transitions in \ce{Ca3Ti2O7} and the performance of the \gls{nep} and \gls{qnep} potentials, heating simulations were conducted for each of the three trained models.
Supercells containing approximately \num{30000} atoms were heated from \qty{1}{\kelvin} to \qty{1500}{\kelvin}, with a heating rate set to \qty{0.4}{\kelvin\per\pico\second}.
For further details on these simulations, as well as the results concerning the two initial models, see the Supplementary Information \cite{SM}. 
This also includes equivalent heating simulations and analysis for \ce{CaTiO3}.

The \gls{qnep} model including target \glspl{bec} was also used when modeling dynamic \gls{dw} motion, with the \gls{pppm} method employed for computational speed.
Compared with regular Ewald summation, \gls{pppm} consistently overestimates the \gls{dw} formation energies, but the difference is only \qty{1.7}{\percent} on average (\autoref{fig:SM_PPPM}).
The time step was set to \qty{1}{\femto\second}, sampling the trajectory every \qty{5}{\pico\second}.
All \gls{dw} supercells were optimized prior to the \gls{md} runs, although an equilibrium configuration was consistently found even with unrelaxed cells. 
All trajectories were visualized with \textsc{ovito} \cite{ovito}. 

\section*{Data availability}
The data that support the findings of this study are openly available in Zenodo at \url{https://doi.org/10.5281/zenodo.22801877} \cite{zenodo}.

\section*{Code availability}
The \gls{nep} and \gls{qnep} models generated in this study, as well as the domain wall structures evaluated, are openly available in Zenodo at \url{https://doi.org/10.5281/zenodo.22801877} \cite{zenodo}, in a format suitable for the \textsc{gpumd} package \cite{fan2022gpumd, XuBuPan25}.
All other software used in this work is publicly available, as described in the Methods section.

\section*{Acknowledgments}
We gratefully acknowledge financial support from the Research Council of Norway (No. 302506), the Swedish Research Council (Nos. 2020-04935 and 2025-03999), and the Knut and Alice Wallenberg Foundation (No. 2024.0042).
Computational resources were provided by UNINETT Sigma2 (Project No. NN9264K), as well as the National Academic Infrastructure for Supercomputing in Sweden (NAISS) at C3SE, partially funded by the Swedish Research Council through grant agreement No. 2022-06725, and the Berzelius resource provided by the Knut and Alice Wallenberg Foundation at NSC.

\section*{Author contributions}
I.C.S. performed the majority of the analysis and wrote the original draft.
E.F. contributed to the analysis, methodology, and supervised the technical work.
L.Ö.W. contributed to the development of the training dataset.
B.A.D.W. and S.M.S. supervised the work.
N.C.B. contributed to overall interpretation and technical discussions.
P.E. led and supervised the project.
I.C.S., S.M.S., and P.E. conceived and designed the study.
All authors reviewed the manuscript and approved the final version.

\section*{Funding}
Open access funding provided by Chalmers University of Technology.

\section*{Competing interests}
The authors declare no competing interests.

\bibliography{references}

\end{document}